\documentstyle[emulateapj]{article}
\newcommand{\kps}{\,\rm{km~s}^{-1}}
\newcommand{\msun}{\,M_{\odot}}
\newcommand{\lsun}{\,L_{\odot}}

\newcommand{\coc}{\,CO(3$\rightarrow$2) }

\begin{document}

\lefthead{Frayer et al.}

\righthead{The $z=2.51$ Submillimeter Galaxy SMM\,J04431+0210}

\submitted{Accepted AJ, \today}

\title{The $z=2.51$ Extremely Red Submillimeter Galaxy SMM\,J04431+0210}

\author{D.\ T.\ Frayer\altaffilmark{1},
L.\ Armus\altaffilmark{1}, 
N.\ Z.\ Scoville\altaffilmark{2},
A.\ W.\ Blain\altaffilmark{2},
N.\ A.\ Reddy\altaffilmark{2},
R.\ J.\ Ivison\altaffilmark{3},~\&
Ian Smail\altaffilmark{4}}

\altaffiltext{1}{SIRTF Science Center, California Institute of Technology
220--06, Pasadena, CA  91125, USA} 
\altaffiltext{2}{Astronomy Department, California Institute of Technology
105--24, Pasadena, CA  91125, USA} 
\altaffiltext{3}{Astronomy Technology Centre, Royal Observatory,
Blackford Hill, Edinburgh EH9 3HJ, UK}
\altaffiltext{4}{Institute for Computational Cosmology, University of
Durham, South Road, Durham, DH1 3LE, UK}

\begin{abstract}

We report the redshift measurement for the submillimeter selected galaxy
SMM\,J04431+0210 (N4) using the Near Infrared Spectrograph on the
Keck\,II telescope.  The data show H$\alpha$, [NII]\,6583,6548, and
[OIII]\,5007 lines at a redshift of $z=2.51$.  The high nuclear
[NII]/H$\alpha$ line ratio is consistent with a LINER or Type-II AGN.
The H$\alpha$ emission is spatially resolved, suggesting the presence of
significant star-forming activity outside the nucleus.  From imaging
with the Near Infrared Camera on the Keck\,I telescope, we find an
extremely red near-infrared color of J$-$K=3.2 for N4.  Follow-up
redshifted \coc observations with the Owens Valley Millimeter Array
constrain the mass of molecular gas to be less than $4\times 10^{10}
\msun$, after correcting for lensing.  The CO to sub-mm flux limit, the
spectroscopic line ratios, and the spectral energy distribution for N4
are all within the range of properties found in other high-redshift
sub-mm sources and local ultraluminous infrared galaxies.  After the
correction for lensing, N4 is the weakest intrinsic sub-mm selected
source with a known redshift and represents the first redshift for the
$<2$\,mJy 850$\mu$m sources which are responsible for the bulk of the
emission from the sub-mm population as a whole.  We argue that N4
contains either an AGN or LINER nucleus surrounded by an extended region
of active star-formation.

\end{abstract}


\section{Introduction}

Since the discovery of the submillimeter population of galaxies (Smail,
Ivison, \& Blain 1997; Hughes et al.\ 1998; Barger et al.\ 1998; Eales
et al.\ 1999), extensive follow-up observations across the entire
electromagnetic spectrum ranging from X-ray to radio wavelengths have
been obtained (see review by Blain et al. [2002]).  Despite these
observational efforts, fundamental questions remain about the population
as a whole, foremost among these are the redshift distribution and
dominant energy source heating the dust.  Detailed studies of individual
sub-mm galaxies are crucial in answering these questions, however,
progress has been slow since many sub-mm sources are extremely faint in
their observed optical bands.  Several of the original potential bright
optical counterparts are now suspected to be mis-identifications based
on deep near-infrared observations (Smail et al. 1999; Frayer et
al. 2000; Dunlop et al. 2003) and redshift estimates based on radio data
(Smail et al. 2000; Webb et al. 2003).

The current data show that the sub-mm population has a mixture of AGN
and starburst characteristics with properties which are roughly
consistent with those of the local population of ultraluminous
($L>10^{12}L_{\odot}$) infrared galaxies (ULIRGs).  The molecular CO
line data for the sub-mm galaxies show the presence of sufficient
molecular gas to fuel high levels of star-formation activity (Frayer et
al. 1998, 1999).  However, the relative importance of AGN and starburst
activity in powering the high luminosities of the sub-mm population is
still open to debate, as is the case with local ULIRGs (e.g., Sanders \&
Mirabel 1996; Genzel et al. 1998; Veilleux, Kim, \& Sanders 1999).  In
order to understand the nature of the sub-mm population, we have been
carrying out multi-wavelength observations of individual systems in the
SCUBA Cluster Lens Survey (Smail et al. 2002).  This survey consists of
15 background sub-mm sources identified from sensitive sub-mm mapping of
seven massive, lensing clusters.  The advantage of this sample is that
the amplification of the background sources allows for deeper source
frame observations.

In this paper we report a redshift of $z=2.5092\pm0.0008$ for the sub-mm
galaxy SMM\,J04431+0210 (N4), obtained from near-infrared observations
with the NIRSPEC instrument on the Keck\,II telescope.  The galaxy N4 at
$04^{\rm h} 43^{\rm m} 07\fs1$, $+02\arcdeg 10\arcmin 25\arcsec$
(J2000) was previously identified as the likely sub-mm counterpart based
on its extremely red I$-$K color (Smail et al. 1999).  An increasing
number of sub-mm sources are now believed to be extremely red objects
(EROs [$R-K>6$]) due to their high dust content (Smail et al. 1999; Gear
et al. 2000; Lutz et al. 2001; Ivison et al. 2001; Wehner, Barger, \&
Kneib 2002).  Since N4 is the brightest K-band (2.2$\mu$m) ERO
counterpart in the Cluster Lens Survey, it is an ideal candidate for
testing whether redshifts can be obtained for the SCUBA population in
the near-infrared.  In addition, N4 has a 850$\mu$m flux density of only
1.6\,mJy, after correcting for lensing, making it one of the
intrinsically weakest sub-mm sources yet identified.  Studies of weak
sub-mm sources such as N4 are crucial to our understanding of the
population as a whole since the majority of the sub-mm background arises
from sources with S(850$\mu$m)$<2$\,mJy (Blain et al. 1999a; Cowie,
Barger, \& Kneib 2002).  We also report on follow-up CO observations
with the OVRO Millimeter Array to search for molecular gas, as well as
optical spectroscopy with the ESI instrument on the Keck\,II telescope
to search for additional rest-frame optical and ultraviolet (UV)
emission lines.  The observations presented in this paper are summarized
in Table~1.

A cosmology of H$_o=70\kps\,{\rm Mpc}^{-1}$, $\Omega_{\rm M}=0.3$, and
$\Omega_{\Lambda}=0.7$ is assumed throughout this paper.  We adopt the
lensing amplification factor of $f=4.4$ which was derived by Smail et
al. (1999) and includes both the amplification due to the foreground
cluster potential and the spiral galaxy near the position of N4.

\section{OBSERVATIONS AND DATA REDUCTION} 

\subsection{Spectroscopy}

N4 was observed with NIRSPEC (Near Infrared Spectrograph) on
Keck\,II\footnote{The W.\ M.\ Keck Observatory is operated as a
scientific partnership among the California Institute of Technology, the
University of California, and the National Aeronautics and Space
Administration.  The Keck Observatory was made possible by the generous
financial support of the W.\ M.\ Keck Foundation.  We are most fortunate
to have the opportunity to conduct observations from the summit of Mauna
Kea which has a very significant cultural role within the indigenous
Hawaiian community.} on 2000 December 26.  NIRSPEC is a spectrograph
operating over 0.95 to 5.4$\mu$m (McLean et al. 1998).  Observations
were obtained in low-resolution mode (R$\sim$1500) with a $0\farcs76$ (4
pixel) wide slit.  The seeing was typically 0$\farcs5$ during the
spectroscopic observations.  The 42$\arcsec$ length slit was placed at a
position angle of $-14^{\circ}$ to contain both N4 and the core of the
bright nearby galaxy $3\farcs1$ from N4 (Fig. 1).  We used the nspec7
filter to search the entire K-band and obtained 130 minutes of useful
on-source integration.  H$\alpha$ emission was detected at $z=2.51$
(Sec.\,3.1).

In 2002 August we followed up the K-band detection of H$\alpha$ by
searching for the [OIII]\,5007,\,4959 doublet and H$\beta$ line in
H-band using NIRSPEC.  We used the same slit setup as done previously
for K-band.  The nspec6 filter was used since it is optimized on the
redward side of H-band at the wavelengths of the redshifted emission
lines.

The NIRSPEC data were combined in pairs at two different positions along
the slit.  This provided a zeroth-order background subtraction as well
as removing the dark and bias levels from the data.  Pixel-to-pixel gain
variations were removed by dividing by a normalized flat produced from
halogen lamp images.  The spectra were transformed to produce a linear
spatial and wavelength scale using standard tasks in the IRAF LONGSLIT
package.\footnote{IRAF is the Image Reduction and Analysis Facility
sofware package written and supported by the National Optical Astronomy
Observatories.}  Observations of a star nodded along the slit provided
spatial calibration, while neon and argon arc lamp lines as well as sky
lines provided wavelength calibration.  A linear background was
subtracted from the frames as a function of wavelength to remove time
variations in the sky emission.  The spectra of bright calibration stars
were used to correct for variations in the detector response as a
function of wavelength.

Given its proximity, residual light from the nearby galaxy is present in
the extracted spectrum of N4 after standard data reduction.  In order to
obtain the true continuum level and spectral slope for N4, the wings of
the galaxy light were subtracted from N4 as a function of wavelength.
There is a natural trade-off between the signal-to-noise (S/N) ratio of
the final spectrum and the wavelength resolution over which the
subtraction is done.  We smoothed over two resolution elements in
wavelength for the galaxy subtraction, which left a slight residual
narrow Paschen-$\alpha$ emission line from the foreground galaxy
($z=0.18$) at $2.1\mu$m.  The resulting continuum level of the K-band
spectrum was calibrated by averaging over the emission from 2 to
2.4$\mu$m and placed on the K-band magnitude scale derived from the NIRC
imaging (Sec.\,2.2).  The absolute flux level of the H-band spectrum was
calculated using the continuum levels of the bright nearby galaxy and N4
at 2.0$\mu$m, where the H- and K-band spectra have overlapping
wavelength coverage.

In 2002 August we also obtained follow-up spectroscopy observations with
the ESI (Echellette Spectrograph and Imager) instrument (Sheinis et
al.\,2002) on Keck\,II to search for additional emission lines in the
rest-frame ultraviolet and optical bands.  The ESI data were taken in
the high-resolution echelle mode which provided wavelength coverage from
3900\AA~ to 1100\AA~ at a resolution of approximately $11\kps$.  A slit
with a width of $1\farcs25$ was used.  The ESI optical spectra were
calibrated using the spectrophotometric standard Feige 110 (Massey et
al.\ 1988).

\subsection{Near-Infrared Imaging}

We observed N4 using the Near Infrared Camera (NIRC) on Keck\,I in 1999
October in K-band and 2001 August in J and K-band.  NIRC employs a
$256\times256$ pixel InSb detector with a pixel scale of $0\farcs15$
(Matthews \& Soifer 1994).  The standard $K$-band filter was used
instead of the bluer $K_{\rm s}$ filter since the object is extremely
red.  We observed using a random dithered pattern with about $10\arcsec$
between adjacent dither positions.  Sixty second integrations were taken
at each position.  A total of 22 and 74\,minutes of on-source data were
obtained in the K and J filters, respectively.  The seeing-disks of the
stars observed throughout the exposures varied from about $0\farcs5$ to
$0\farcs7$ (FWHM).

Sets of dark frames from each night were subtracted from each exposure
to remove the dark current as well as the bias level.  The
dark-subtracted exposures were divided by a normalized skyflat.  Frames
were sky-subtracted using temporally--adjacent images.  Individual
exposures were aligned to the nearest pixel using common objects in the
frames.  The data were placed on the Vega-magnitude scale from
observations of a set of near-infrared standard stars (Persson et al.\
1998) taken at a range of air masses to correct for atmospheric
extinction.  Based on the dispersion in the zero-points derived
throughout the nights, the uncertainty of the derived magnitude scale is
estimated to be better than 0.06 magnitude.

\subsection{Millimeter Interferometry}

Based on the redshift derived from the NIRSPEC observations, N4 was
observed using the Owens Valley Millimeter Array\footnote{The Owens
Valley Millimeter Array is a radio telescope facility operated by the
California Institute of Technology and is supported by NSF grant AST
9981546.} to search for CO emission between 2001 January and April.  A
total of 23 hours of usable integration time on source was obtained in
two configurations of six 10.4m telescopes.  The array configurations
yielded a $6\arcsec \times 5\arcsec$ synthesized beam adopting
natural-weighting.  We used a digital correlator configured with
$112\times4$~MHz channels centered on 98.5174 GHz to search for \coc
emission at the corresponding H$\alpha$ line redshift of $z=2.510$.  The
spectral-line bandwidth corresponds to 1400$\kps$ ($\Delta z = \pm
0.005$), which is sufficiently large to observe the entire H$\alpha$
line width, including the redshift uncertainty (Sec.\,3.1).  In addition
to the CO line data, 3mm continuum data were recorded with a bandwidth
of 2~GHz.  The nearby quasar J0433+053 was observed every 25 minutes for
gain and phase calibration.  Absolute flux calibration was determined
from observations of the standard calibrator 3C\,273 whose flux history
is monitored by observations of Uranus and Neptune.  The absolute flux
calibration uncertainty for the data is approximately 15\%.  The data
were reduced using the OVRO MMA software (Scoville et al.\ 1983) and
standard tasks in AIPS.\footnote{AIPS is the Astronomical Image
Processing System sofware package written and supported by the National
Radio Astronomy Observatory.}

\section{RESULTS}

The identification of N4 as the sub-mm counterpart is fairly secure
based on previous studies.  As discussed by Smail et al. (1999), the
nearby $z=0.18$ spiral galaxy can be reliably ruled out based on its
radio/sub-mm flux density ratio and its 450$\mu$m limit.  The random
probability of finding a bright ERO similarly as red ($R-K>6$) as N4
within the $3\arcsec$ positional uncertainty of the SCUBA data is less
than 0.1\%, based on the observed ERO surface densities for $K<20$
(Thompson et al. 1999; Smith et al. 2002).  If we relax the sub-mm
positional uncertainty and account for lensing, the likelihood of a
chance coincidence with an ERO is still less than 1\%, and no other
candidate red objects are nearby.  The identification of N4 as the
sub-mm counterpart has recently been confirmed by the detection of CO
emission from N4 at the H$\alpha$ redshift with the IRAM PdB
interferometer (R. Neri et al. 2003, in preparation).  The OVRO CO
upper-limit presented in this paper is consistent with the more
sensitive PdB observations.

\subsection{Spectroscopy}

An emission-line complex is clearly detected in the K-band spectrum of
N4 (Fig.~2), which is identified as H$\alpha+$[NII] at $z=2.51$.  We
also detect evidence for a weak emission line at 1.756$\mu$m, which
corresponds to [OIII]\,5007\AA~ at the H$\alpha$ redshift (Fig.~3).
Fits to the lines and continuum were made using the SPECFIT package in
IRAF (Table~2).  In fitting the H$\alpha+$[NII] line complex, we have
allowed the widths and fluxes of the H$\alpha$ and [NII]6583\AA~ lines
to be free, while fixing the [NII]6548\AA~ line to have a flux of 30\%
that of [NII]6583\AA.  Both [NII] lines were constrained to have the
same FWHM.  The continuum was fitted with a linear function.  Initially
the position of the [NII]6583\AA~ line was fixed with respect to
H$\alpha$, but this requirement was later relaxed to derive the redshift
and to confirm the identification.  From the fit to the total K-band
spectrum, we derive a redshift of $z=2.5092\pm0.0008$ based on the
H$\alpha$ and NII\,6583 lines.  The redshift uncertainty includes the
error from the fit and the systematic uncertainty associated with the
derived wavelength scale.  The intrinsic H$\alpha$ line width is
$520\pm40\kps$, corrected for instrumental resolution, while the [NII]
lines are slightly narrower, having an intrinsic FWHM of $440\pm60\kps$.
The [NII]\,6583/H$\alpha$ line flux ratio is found to be $0.47\pm0.06$.
No broad-line H$\alpha$ component is readily visible in the data.  The
1$\sigma$ upper-limit on a $5000\kps$ broad H$\alpha$ line is about 10\%
of the H$\alpha$ narrow-line flux.

The H-band data were fitted adopting the same techniques as used for the
K-band data.  The [OIII] line was fitted to derive the wavelength
position, line width, and line flux.  In deriving the H$\beta$ limit, we
assumed the H$\alpha$ redshift and linewidth.  The observed limit on the
[OIII]5007/H$\beta$ flux ratio is greater than 1.5 ($3\sigma$).

The observed Balmer decrement of H$\alpha$/H$\beta > 6$ ($3\sigma$)
implies significant extinction.  Assuming the standard intrinsic
H$\alpha$/H$\beta$ value of 2.85 (Case-B Balmer recombination decrement
for T$=10^4$\,K and N$_e=10^{4}$\,cm$^{-3}$ [Osterbrock 1989]) and the
Galactic reddening function parameterized by Lequeux et al. (1979), we
derive an extinction at H$\alpha$ of greater than $1.6$ mag and a color
excess of E(B-V) $>0.7$ mag.  The high-level of extinction implied by
the H$\alpha$/H$\beta$ ratio is consistent with the non-detection of
Ly$\alpha$ emission from the ESI observations (Table 2).

Figure 4 shows the two-dimensional image of the K-band spectrum.  The
H$\alpha$ emission is spatially resolved, while the [NII] emission is
consistent with a compact object.  The size of H$\alpha$ emission region
is about 1.1\,arcsec, after deconvolution with a Gaussian representing
the seeing point spread function.  Corrected for lensing, the intrinsic
diameter of the detected H$\alpha$ emission regions is about 2\,kpc.  If
the observed velocity gradient is due to rotation, a rough estimate of
the dynamical mass can be made based on the H$\alpha$ line width.  The
H$\alpha$ emission is spread over 750$\kps$ (FWZI), and using a radius
of 1.0\,kpc we calculate a dynamical mass of $3\times 10^{10}\, {\rm
sin}(i)^{-2} \msun$, where the inclination ($i$) is unknown.  The
estimated dynamical mass of N4 is comparable with masses found in the
central regions of normal galaxies in the local universe.

Figure 5 shows spectra extracted from the different regions of N4.  The
[NII]\,6583/H$\alpha$ line flux ratio varies measurably as a function of
distance from the nucleus of N4.  The off-nuclear data have a
[NII]\,6583/H$\alpha$ line flux ratio of $0.16\pm0.06$, consistent with
star-forming regions, while the nuclear region has a much higher
ratio of $0.66\pm0.09$, indicative of an AGN or a LINER (Osterbrock
1989).  Therefore, the data for N4 are consistent with a narrow-line AGN
surrounded by an extended region of active star-formation (Baldwin,
Phillips, \& Terlevich 1981; Veilleux \& Osterbrock 1987) or a powerful
starburst with large-scale nuclear shocks driven by a nuclear superwind.
(Armus, Heckman, \& Miley 1989; Heckman, Armus, \& Miley 1990) .

\subsection{Imaging}

The sensitive {\it HST} R-band (F702W filter) and Keck LRIS I-band
images of N4 have been previously discussed by Smail et al. (1999).  We
compare these optical images with the deep J- and K-band data presented
in this paper.  In all bands we subtracted the nearby galaxy using the
IRAF task ELLIPSE, which enabled source masking while successfully
removing the wings of the galaxy emission.  The images were aligned
using common sources in the frames.  For optimization in K-band, we
adopted a $2\arcsec$ diameter aperture for photometry.  Although the S/N
ratios from the aperture measurements in the other bands are lower than
for K-band (Table~3), the emission peak was detected in each of the four
bands with a S/N ratio greater than five.  The errors on the
measurements represent the rms of multiple aperture measurements on the
sky around N4, avoiding the galaxy residuals from the central disk.  We
measure the colors of $R-K= 6.34\pm0.27$ mag and $J-K= 3.15\pm0.29$ mag
for N4.  The $R-K$ measurement presented here is slightly bluer than the
original $R-K>6.7$ color quoted by Smail et al. (1999), which is
attributed to the difference in the adopted aperture size and methods of
galaxy subtraction.

We have resolved N4 in both the R and K-band images (Fig. 6).  After
deconvolution of the $0\farcs6$ stellar PSF, approximated by a Gaussian,
we derive a FWHM size of $1\farcs1\pm0\farcs2$ for the K-band emission,
consistent with the H$\alpha$ source size (Sec.\,3.1).  The K-band image
is extended in the north-south direction, along the expected direction
of lensing amplification, while the R-band image appears extended
east-west.  The K-band peak is marginally offset along the line of
R-band elongation, but its derived position is coincident with the
R-band peak within the errors.  Offsets between the K-band and R-band
emission regions are not unexpected given the high level of extinction
within N4 and given previous results for sub-mm galaxies (Ivison et
al. 2001).

\subsection{Dust and Molecular Gas}

The OVRO CO observations yield a 3$\sigma$ upper-limit of 4.5 mJy/beam,
averaging over the H$\alpha$ line-width of 520$\kps$ with a
natural-weighted beam size of $6\arcsec \times 5\arcsec$.  The limit on
the integrated \coc flux density of $S(CO)<2.5 \,{\rm Jy}\kps$
corresponds to a CO luminosity of $L^{\prime}(\rm{CO})<1.9\times
10^{10}\,{\rm K}\kps\,{\rm pc}^{2}$, correcting for lensing.  The CO to
H$_2$ conversion factor is still uncertain for the sub-mm population.
Adopting a conversion factor of $L^{\prime}(\rm{CO})/{\rm
M(gas)}=2\,{\rm K}\kps\,{\rm pc}^{2}(\msun {\rm yr}^{-1})^{-1}$, which
is appropriate for low redshift ULIRGs (Solomon et al. 1997; Scoville,
Yun, \& Bryant 1999), the observations imply M(gas)$< 4 \times
10^{10}\msun$ for N4.  The limit on the gas mass is comparable to the
total dynamical mass estimated from the H$\alpha$ line-width
(Sec.\,3.1).  The \coc luminosity limit is slightly lower than the
previous CO detections for sub-mm selected galaxies (Frayer et
al. 1998, 1999), after converting to the same cosmology, but is well
within expectations given the lower $850\mu$m flux density for N4.

We failed to detect 3mm continuum emission from N4, achieving a
3$\sigma$ upper-limit of S(100\,GHz)$<1.5$\,mJy.  The 3mm continuum
limit is consistent with the previous SCUBA measurements of N4 (Smail et
al. 1999).  The spectral energy distribution (SED) for N4 is consistent
with those found for previous sub-mm galaxies (Ivison et al.\ 2000) and
local ULIRGs.  Based on its redshift and the limit on its
850$\mu$m/1.4GHz flux ratio, we estimate a dust temperature of T$>40$\,K
from the Carilli \& Yun (1999) relationship plotted as a function of
temperature by Blain et al. (2002).  The dust temperature is consistent
with both cool and warm infrared SEDs found for ULIRGs (Soifer et
al. 1989).

\section{DISCUSSION}

As mentioned in Section 3, the rest-frame optical emission-line flux
ratios of N4 imply a composite source.  The relatively large nuclear
[NII]/H$\alpha$ and [OIII]/H$\beta$ ratios are suggestive of a
non-thermal ionizing source, although we cannot rule out shock
excitation from a wind.  The spatially-resolved H$\alpha$ line emission
and the lower off-nuclear [NII]/H$\alpha$ ratio suggest a change to hot
stars as the dominant source of ionizing photons outside the nucleus.

The observed [NII]\,6583/H$\alpha$ and [OIII]\,5007/H$\beta$ line ratios
for N4 are consistent with the average values found in low redshift
ULIRGs with L$_{\rm ir} > 2\times 10^{12} \lsun$, albeit there is a
large range of ratios found for local ULIRGs (Veilleux et al. 1999).
The limits on the [OI]\,6300 and [SII]\,$\lambda\lambda$6716,6731 lines
do not provide any additional constraints favoring one classification
over another.  Unfortunately, the [OI]\,6300 line for N4 is redshifted
to the wavelength of the Pa$\alpha$ from the foreground galaxy.
Although there is tentative evidence for [OI]\,6300 emission even after
careful galaxy subtraction, no robust measurement could be derived for
this line.  The H$\alpha$ equilvalent width (EQW) and relatively narrow
line width suggest that N4 does not contain a broad-line Type-I AGN, or
that the broad-line region is heavily obscured.  A higher
S/N spectrum would be required to place more stringent
limits on the H$\alpha$ broad-line flux from N4.

The only X-ray data that currently exists for N4 are from {\it ROSAT},
and these data are not sensitive enough to constrain the fraction of the
bolometric luminosity which may arise from an AGN.  Given the unknown
contribution from a possible AGN and the uncertainty in the level of
extinction, the star-formation rate (SFR) is fairly uncertain for N4.
We can derive upper and lower bounds on the SFR based on the H$\alpha$
and sub-mm emission.  The off-nuclear H$\alpha$ line luminosity
uncorrected for extinction provides a lower-limit on the SFR in N4.  At
the observed resolution, approximately 40\% of the total H$\alpha$
emission is outside the nucleus.  This off-nuclear H$\alpha$ emission
corresponds to a line luminosity of L(H$\alpha) = 2\times10^{8} \lsun$,
corrected for lensing.  Adopting a standard relationship between the
H$\alpha$ line luminosity and the SFR (Kennicutt 1983), we find that SFR
in N4 is greater than $6 \msun\,{\rm yr}^{-1}$, uncorrected for
extinction.  Unfortunately, the H$\beta$ limit is not sensitive enough
to provide a useful estimate of the extinction level outside the
nucleus.  For comparison, the total H$\alpha$ line luminosity corrected
for the extinction found from the limit on the Balmer decrement would
imply a total SFR of $>70 \msun\,{\rm yr}^{-1}$.

A upper-limit to the SFR can be derived by assuming that all the
far-infrared (FIR) radiation is due to star-formation, neglecting any
contribution due to a possible AGN.  The FIR luminosity for N4 is
derived assuming that its SED is somewhere between that of the
infrared-warm ULIRG Mrk\,231 and the infrared-cool ULIRG Arp\,220.
Placing these ULIRGs at the redshift of N4 and scaling to the de-lensed
850$\mu$m flux of N4, we estimate an intrinsic FIR luminosity of L(FIR)$
=$2--5$\times 10^{12}\lsun$ for N4.  This FIR luminosity corresponds to
a SFR of massive stars of SFR(M$>5\msun) \simeq$200--500$\msun\,{\rm
yr}^{-1}$, using the relationship given by Condon (1992).  Including the
presence of lower mass stars with the IMF adopted by Kennicutt (1983),
we obtain a total SFR$\simeq$800--2000$\msun\,{\rm yr}^{-1}$.  Hence,
from the H$\alpha$ and sub-mm emission, the star-formation rate
of N4 is probably at least $100\msun\,{\rm yr}^{-1}$ and could be as
large as $2000 \msun\,{\rm yr}^{-1}$.

Despite the high level of obscuration inferred for the sub-mm galaxies,
the recent Ly$\alpha$ detections of radio-selected sub-mm sources by
Chapman et al. (2003) suggest that significant rest-frame UV light can
escape from the sub-mm sources.  For N4 we failed to detect Ly$\alpha$
emission.  The sensitivity of the ESI observations would have been
sufficient to detect many of the Ly$\alpha$ sources seen by Chapman et
al. (2003).  However, given the fact that the Ly$\alpha$/H$\alpha$ ratio
can span three orders of magnitudes for ULIRGs ($<0.01$ -- 10; {\it HST}
STIS observations by Surace et al. 2003, in preparation), the observed
limit of Ly$\alpha$/H$\alpha < 0.14$ for N4 is not particularly
constraining.

N4 has a composite spectrum consisting of a narrow-line AGN/LINER
nucleus surrounded by a resolved starburst.  Other well-studied sub-mm
galaxies have shown a wide range of characteristics.  The system
SMM\,J02399-0126 at $z=2.8$ shows an AGN spectrum (Ivison et al.\ 1998)
which has been classified as broad absorption line QSO (Vernet \&
Cimatti 2001), while SMM\,J14011+0252 at $z=2.6$ shows no evidence of an
AGN and is thought to be dominated by star formation (Barger et al.\
1999; Ivison et al.\ 2000).  The most distant know sub-mm galaxy
SMM\,J09431+4700 ($z=3.3$) contains a narrow-line Seyfert-1 AGN (Ledlow
et al. 2002), and the sub-mm galaxy N2\,850.4 ($z=2.4$) is a composite
source (AGN+starburst) showing strong outflows and P-Cygni profiles
indicative of stellar winds (Smail et al.\ 2003).  All of these sub-mm
galaxies are significantly more luminous than N4 and can be classified
as hyperluminous sources ($L\ga 10^{13} \lsun$).  It may not be
surprising that most these of the hyperluminous sub-mm sources contain
AGN given that this is seen at lower redshifts (e.g., Evans et
al. 1998).  The results for N4 suggest that even the lower luminosity
sub-mm sources could also be partially powered by AGN.

Although there is evidence for AGN in many sub-mm sources, the bulk of
the infrared light for the high-redshift sub-mm population is still
thought to be dominated by star-formation (Blain et al.\ 1999b, 2002).
The X-ray data support this conclusion (Barger et al.\ 2001; Alexander
et al.\ 2003; Almaini et al. 2003).  The deepest X-ray data from
Alexander et al.\ (2003) suggest that even though a significant faction
of the bright sub-mm sources contain an AGN, the AGN typically have low
luminosities and contribute negligibly to the total bolometric
luminosity of the population.

In the local universe, the apparent fraction of AGN dominated ULIRGs
increases at luminosities above L$_{\rm ir} > 2\times 10^{12} \lsun$
(Veilleux et al. 1999).  It is unclear if the high-redshift sub-mm
population of galaxies will follow this trend.  The relative importance
of AGN and star-formation activity in the sub-mm population will be
constrained with future sensitive X-ray observations, {\it SIRTF}
observations that will distinguish between infrared warm (AGN) versus
cool (starburst) sub-mm galaxies, and high-resolution mm and CO
interferometric observations that will constrain gas masses and source
sizes.

\section{CONCLUSIONS}

We report the redshift of $z=2.51$ for the counterpart of the sub-mm
galaxy N4.  We have detected the H$\alpha$, [NII], and [OIII] lines and
present upper-limits for Ly$\alpha$, H$\beta$, and \coc lines.  The line
ratios and observed SED are consistent with the range of properties
found in local ULIRGs.  The data show that N4 is comprised of a nuclear
region showing a LINER/Type-II AGN spectrum with an extended H$\alpha$
component, presumably arising from an extended starburst.  It is still
unclear whether N4 is predominately powered by AGN or star-forming
activity.  Future multi-wavelength observations should provide better
constraints on the source of its immense luminosity.

\acknowledgements

We thank the staff at Keck Observatory and the Owens Valley Millimeter
Array who have made these observations possible.  We thank B.\ T.\
Soifer and E.\ Egami for obtaining an early H-band spectrum of N4.  DTF
and LA are supported by the Jet Propulsion Laboratory, California
Institute of Technology, under contract with NASA.  AWB acknowledges
support from the NSF under grant AST-0205937. NAR acknowledges support
from a NSF Graduate Research Fellowship, and IS acknowledges support
from the Royal Society and the Leverhulme Trust.

%table1
\begin{deluxetable}{llccr}
\tablenum{1}
\tablecaption{Observational Log}
\tablewidth{400pt}
\tablehead{\colhead{UT Date(s)}&\colhead{Telescope}
&\colhead{Instrument}&\colhead{Line, Continuum}
&\colhead{Integration Time}}
              
\startdata

1999\,Oct\,01 & Keck\,I & NIRC & 2.2$\mu$m & 10 min \nl
2000\,Dec\,26 & Keck\,II & NIRSPEC & H$\alpha$ & 130 min \nl
2001\,Jan-May & OVRO & 3mm SIS Mixer & \coc, 3mm & 1380 min \nl
2001\,Aug\,29 & Keck\,I & NIRC & 2.2$\mu$m & 12 min \nl
2001\,Aug\,29-31 & Keck\,I & NIRC & 1.2$\mu$m & 74 min \nl
2002\,Aug\,08-09 & Keck\,II & ESI & Ly$\alpha$ & 110 min \nl
2002\,Aug\,16    & Keck\,II & NIRSPEC & [OIII],H$\beta$ & 70 min\nl

\enddata 
\end{deluxetable}

%table2
\begin{deluxetable}{lcccc}
\tablenum{2}
\tablecaption{Spectroscopic Properties of SMM\,J04431+0210\tablenotemark{a}}
%\tablewidth{350pt}
\tablehead{\colhead{Line}&\colhead{$\lambda_{\rm
obs}$}&\colhead{FWHM}\tablenotemark{b}&\colhead{EQW}\tablenotemark{c}&
\colhead{Flux}\nl \colhead{
}&\colhead{(\AA)}&\colhead{$(\kps)$}&\colhead{(\AA$_{\rm
rest}$)}&\colhead{($10^{-17}$\,erg\,s$^{-1}$\,cm$^{-2}$)}}
              
\startdata

H$\alpha$  & 23030$\pm$5 & 520$\pm$40& 66$\pm$4& 15.8$\pm$1.0\nl
[NII]\,6583  & 23102$\pm$5 & 440$\pm$60& 32$\pm$4& 7.4$\pm$0.9\nl
[OIII]\,5007 & 17560$\pm$8 & 580$\pm$160& 41$\pm$9&
4.0$\pm$0.8\nl 
H$\beta$     & ...         & ...   & ...  & $< 2.7
\,(3\sigma)$\tablenotemark{d} \nl  
L$\alpha$  & ...         & ...   & ...  & $< 2.2
\,(3\sigma)$\tablenotemark{d} \nl 
\coc &...&...&...& $< 2.5\,{\rm Jy}\kps$ ($3\sigma$)\tablenotemark{d}\nl
\enddata 
\tablenotetext{a}{Values calculated from fits to the unsmoothed data,
uncorrected for lensing or extinction.}
\tablenotetext{b}{Intrinsic FWHM corrected for instrumental resolution.}
\tablenotetext{c}{Rest-frame EQW.}  
\tablenotetext{d}{All limits assume an intrinsic $520\kps$ FWHM line
width found for the H$\alpha$ line.}

\end{deluxetable}

%table3
\begin{deluxetable}{lcccl}
\tablenum{3}
\tablecaption{Photometric Properties of SMM\,J04431+0210}
\tablewidth{350pt}
\tablehead{\colhead{Observed Band\tablenotemark{a}}&\colhead{Measured Value}
&\colhead{Notes}}
              
\startdata

R        & 25.75$\pm$0.25 & {\it HST} (F702W) \nl
I        & 26.5$\pm$0.6   & Keck LRIS  \nl
J        & 22.56$\pm$0.28 & Keck NIRC  \nl
K        & 19.41$\pm$0.09 & Keck NIRC  \nl
850$\mu$m  &  7.2$\pm$1.7\,mJy   & Smail et al.\ (1999) \nl
450$\mu$m  & $< 60$\,mJy ($3\sigma$) & Smail et al.\ (1999) \nl
3\,mm      & $< 1.5$\,mJy ($3\sigma$) & OVRO \nl
20\,cm     & $< 70$\,$\mu$Jy ($3\sigma$) & Smail et al.\ (1999) \nl

\enddata 
\tablenotetext{a}{All magnitudes are on the Vega scale, are measured with
a 2$\arcsec$ diameter aperture, and are uncorrected for lensing.}
\end{deluxetable}

\newpage
%fig1
\begin{figure}[t]
\includegraphics{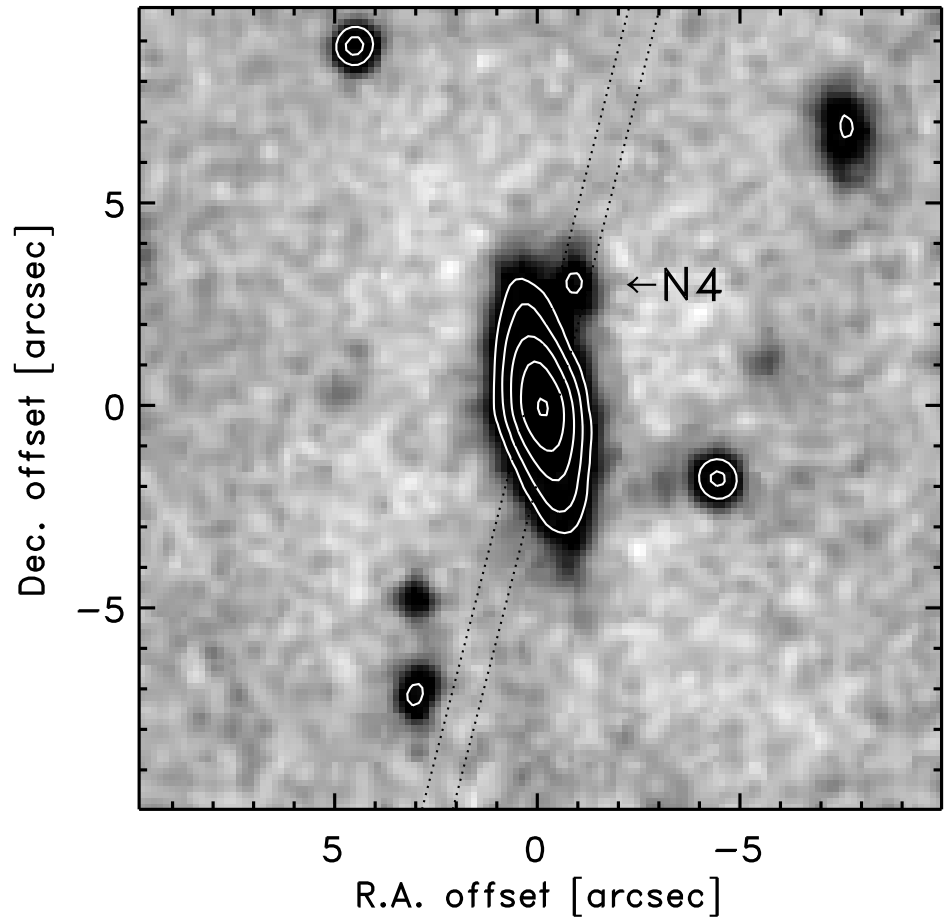}
\vspace*{5.0in}
\caption{The Keck K-band ($2.2\mu$m) image of N4.  The slit location
for the NIRSPEC spectroscopic observations is shown by the dotted lines.
The grey-scale is plotted on a logarithmic scale, and the contour levels
are separated by a factor of two in flux density, starting at the
$15\sigma$ level.}
\end{figure}

\newpage
%fig2
\begin{figure}[t]
\includegraphics{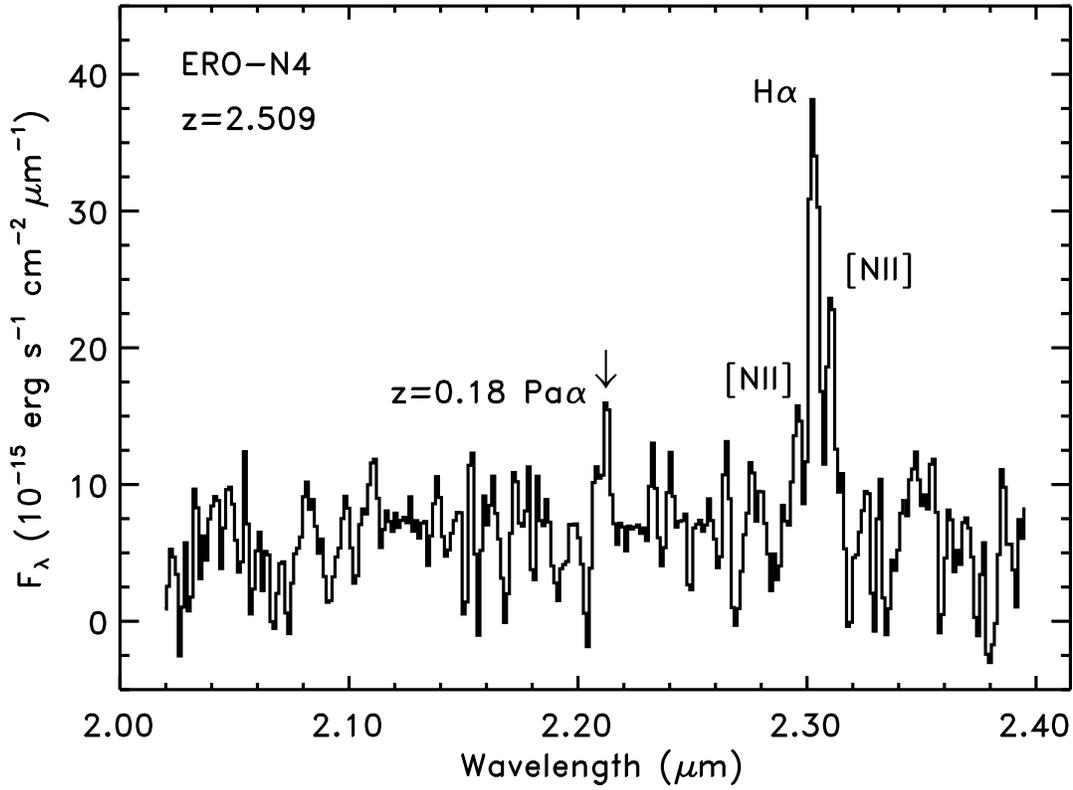}
\vspace*{5.0in}
\caption{The observed K-band spectrum of N4.  The H$\alpha$ and
[NII] lines place N4 at a redshift of $z=2.5092$.  The spectrum has
been calibrated based on the K-band continuum level of N4 and has been
smoothed by 26\AA.  A small amount of residual Pa$\alpha$ emission from
the foreground galaxy ($z=0.18$) remains after the subtraction of the
wings of its continuum from the N4 data.}
\end{figure}

\newpage
\begin{figure}[t]
%fig3
\includegraphics{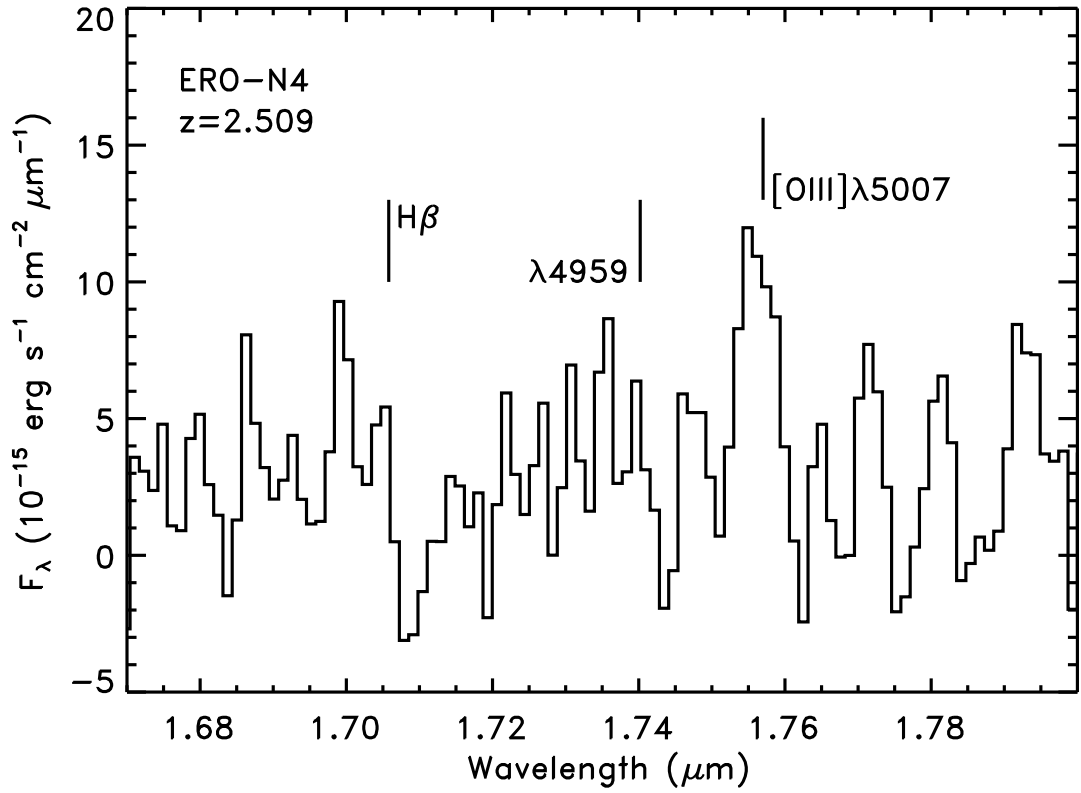}
\vspace*{5.0in}

\caption{The observed H-band spectrum of N4.  The [OIII]\,5007 line is
detected, and only upper-limits can be derived at the location of the
redshifted H$\beta$ and [OIII]\,4959 lines. The data have been smoothed
by 25\AA. }

\end{figure}

\newpage
%fig4
\begin{figure}[t]
\includegraphics{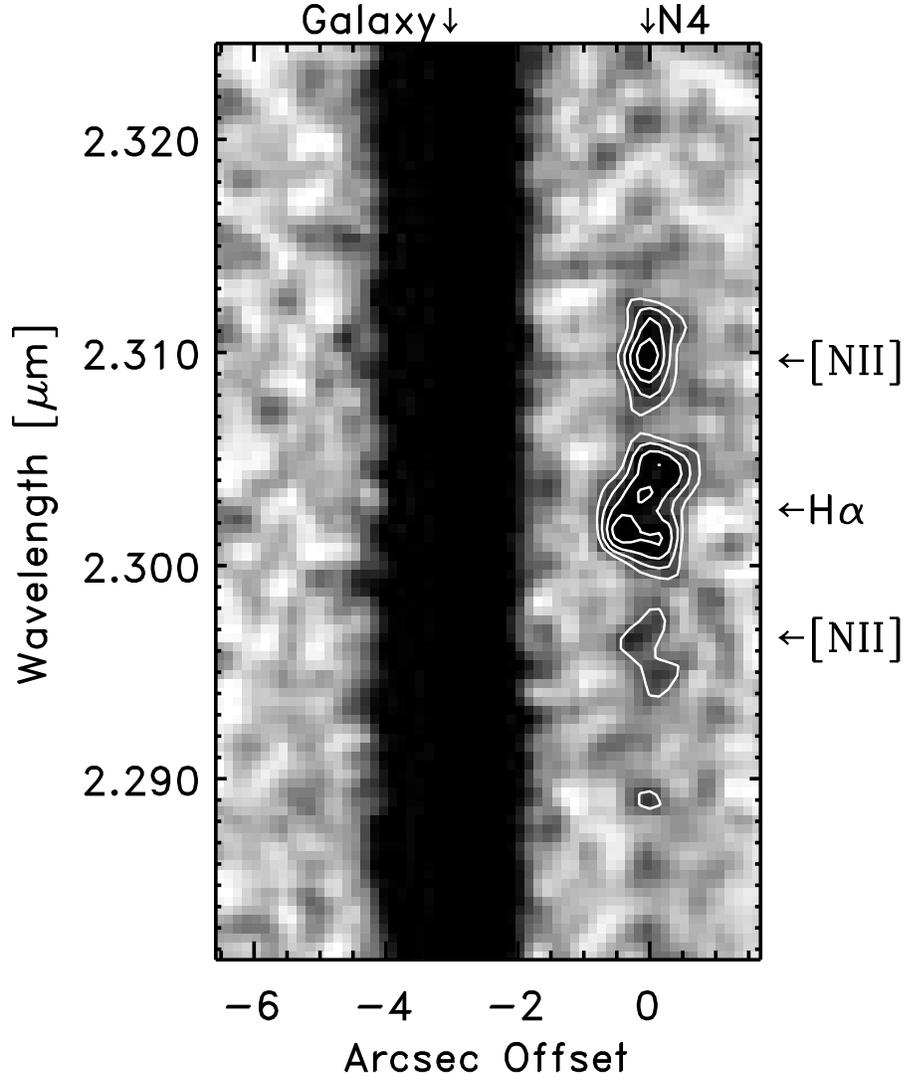}
\vspace*{5.0in}

\caption{The image of the K-band spectrum of N4.  The H$\alpha$
emission is spatially extended unlike the spatially unresolved
[NII]\,6583 line.  The extended H$\alpha$ emission along with the
observed line ratios suggests the association of bright star-forming
regions around a central AGN and/or LINER nucleus.  The contours for N4
start at $5\sigma$ and are incremented by $2\sigma$.}

\end{figure}

\newpage
%fig5
\begin{figure}[t]
\includegraphics{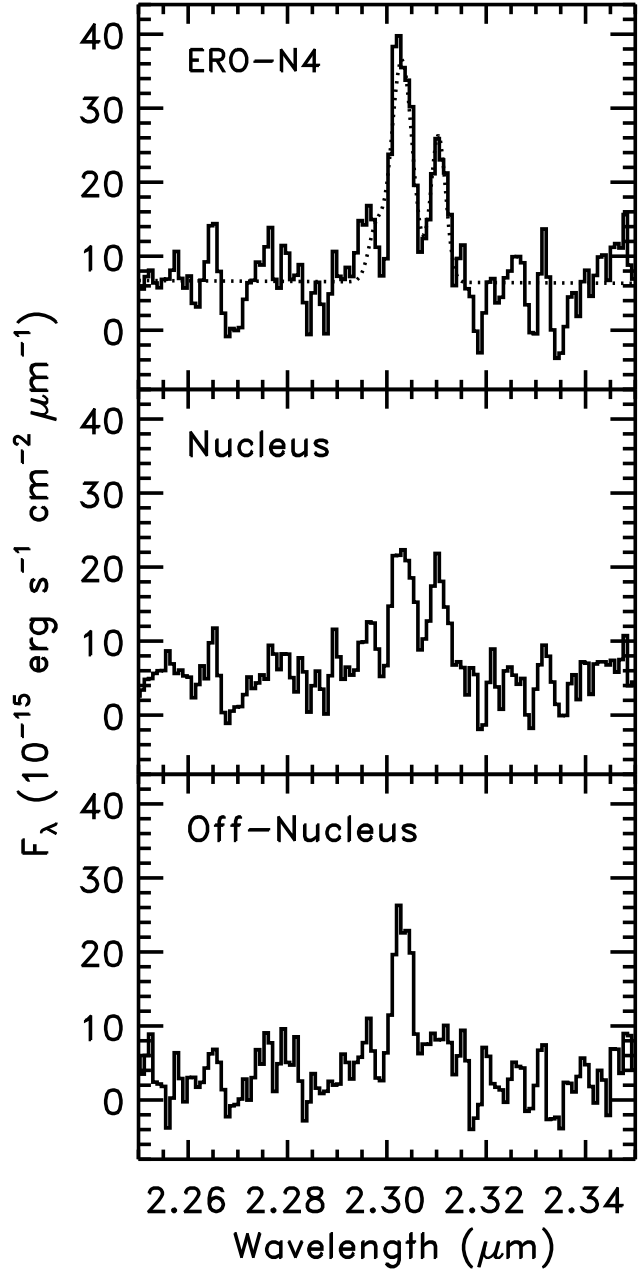}
\vspace*{7.0in}

\caption{The K-band spectra of N4 extracted over different spatial
regions smooth over 17\AA.  The top spectrum shows the total spectrum
over all emission regions.  The middle spectrum shows the central
nucleus, while the bottom spectrum represents off-nucleus emission.  The
variation of the H$\alpha$/[NII] ratio suggests that N4 is a composite
system with active star formation off-nucleus.  The dotted-line in the
top spectrum shows the best fit model to the unsmoothed data for the
H$\alpha+$[NII] complex discussed in Sec.~(3.1).}
\end{figure}

\newpage
%fig6
\begin{figure}[t]
\includegraphics{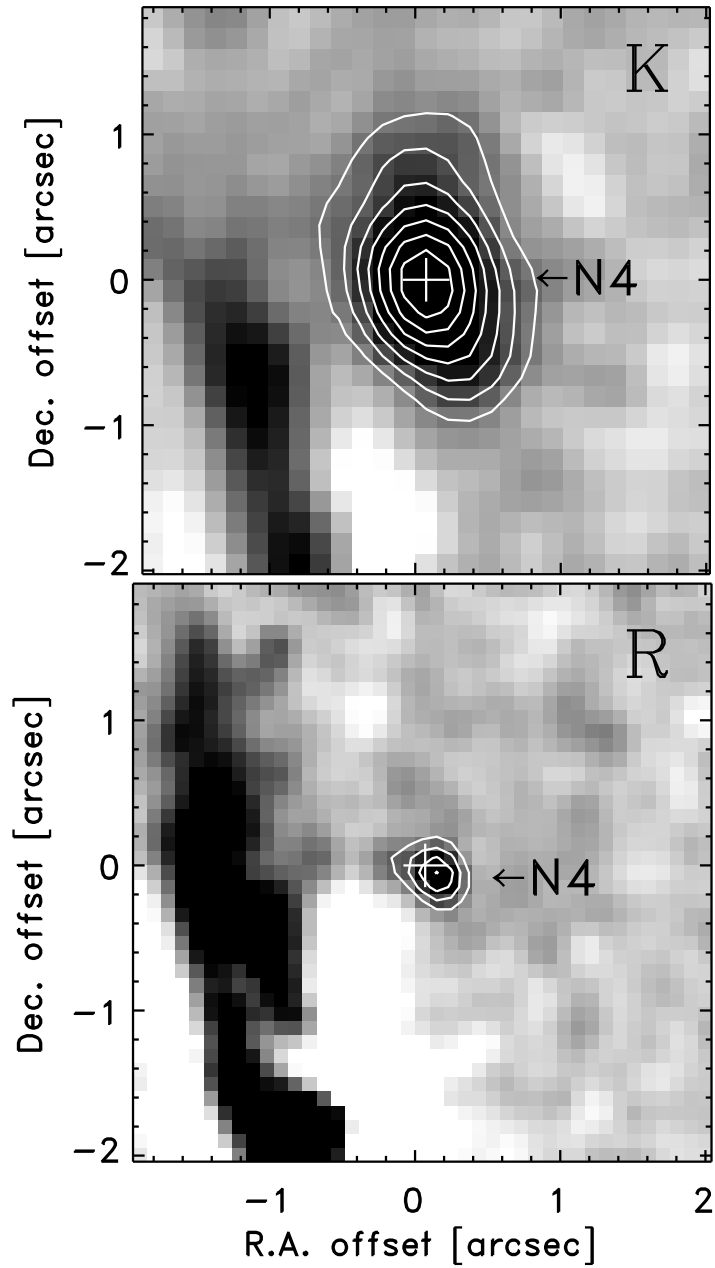}
\vspace*{7.0in}

\caption{The K- and R-band images of N4 after the subtraction of the
wings of the bright foreground galaxy.  The cross shows the position of
the K-band peak, and its size represents the uncertainty in registering
the K- and R-band data.  The logarithmic grey-scale is plotted from
$-3\sigma$ to $+10\sigma$, and the contour levels start at $4\sigma$ and
are incremented by $2\sigma$.  Although the wings of the galaxy were
removed, the alternating white and dark regions in the lower-left of the
images are due to the over and under subtraction of the central disk,
respectively.}

\end{figure}

\end{document}